\documentclass[preprint2]{aastex}
\usepackage{amssymb}
\usepackage{epsfig}
\usepackage{natbib}

\begin{document}
\title{A model for the non-universal power-law of the solar wind sub-ion scale magnetic spectrum}
\shorttitle{Sub-ion scale solar wind magnetic spectrum}

\author{T. Passot and P.L. Sulem}

\affil{Laboratoire Lagrange,\\
Universit\'e C\^ote d'Azur, CNRS,
Observatoire de la C\^ote d'Azur}
\affil{CS 34229, 06304 Nice Cedex 4, France}
\email{passot@oca.eu; sulem@oca.eu}


\begin{abstract}
A phenomenological turbulence model for kinetic Alfv\'en waves in a magnetized
collisionless plasma, able to reproduce the non-universal
power-law spectra observed at the sub-ion scales
in the solar wind and the terrestrial magnetosphere, is presented.
The process of temperature homogenization along distorted magnetic field lines, 
induced by Landau damping, affects the turbulence
transfer time and results in a steepening of the sub-ion power-law spectrum
of critically-balanced turbulence, whose exponent is 
sensitive to the ratio between the Alfv\'en wave period and the nonlinear 
timescale. Transition from large-scale
weak turbulence to smaller scale strong turbulence is captured
and non local interactions, relevant in the case of steep spectra, are accounted for.

\end{abstract}




\section{Introduction}
Spacecraft measurements both in the solar wind and the Earth magnetosphere 
\citep{BC13,ACS13,ALM08} show power-law
energy spectra for magnetic turbulent fluctuations. At MHD scales, where kinetic
effects are subdominant, observations support 
an Alfvenic energy cascade where the magnetic fluctuations transverse to the ambient field display a
spectrum close to the $k_\perp^{-5/3}$ prediction based on a ``critical balance'' \citep{GS95,NS11} 
between the characteristic times of the nonlinear transverse dynamics and 
of the Alfv\'en wave propagation
along the magnetic field lines. At sub-ionic scales,  a
power law is also observed in a range extending from the proton ($\rho_i$) 
to the electron ($\rho_e$)
Larmor radius \citep{Sah09,Sah10,Sah11,Alexandrova12,Chen13}, but 
the exponent appears to be less universal, with a distribution peaked near $-2.8$ 
and covering the interval $[-3.1,-2.5]$ (Fig. 5 of  \citet{Sah13}). 
For comparison, the magnetic spectral exponent in the MHD range is estimated
as $-1.63\pm 0.14$ by \citet{Smith06}.

Gyrokinetic simulations at $\beta=1$ display 
a sub-ion power-law spectrum with comparable exponents ($-2.8$ in 
\citet{Howes11b} or $-3.1$ in \citet{Told15}). Fully kinetic 
particle-in-cell (PIC) codes with  a reduced mass ratio  \citep{Wan15}, 
as well as hybrid Eulerian Vlasov-Maxwell models \citep{Servidio15} also predict
steep spectra at small scales, associated with coherent structures and deformation of the 
particle distribution functions. 

At sub-ion scales, two types of waves play a dynamical role: whistlers
for which ions are approximately cold and kinetic compressibility is negligible,
and  (low-frequency) kinetic
Alfv\'en waves (KAWs) for which density fluctuations are significant. 
Reduced fluid-like models for the nonlinear dynamics
of such waves have been developed 
and numerically simulated, leading to a $-8/3$ sub-ion spectral 
exponent  \citep{Scheko09,Boldyrev12,Boldyrev13,Meyrand13}.
Nevertheless a comprehensive understanding of turbulence at these scales
 is still missing.
Carrying  the critical-balance phenomenology to the KAW cascade
leads to a $k_\perp^{-7/3}$  energy spectrum, 
significantly shallower than observed. Still assuming critical balance, 
\citet{Boldyrev13} proposed  a steeper spectrum resulting from
coherent structures and intermittency corrections. Differently, 
\citet{Howes08} suggested a balance between Landau damping and energy transfer. 
The $k_\perp^{-7/3}$ spectrum is then 
multiplied by an exponential factor originating from 
the variation of the energy flux along the cascade. This model was extended 
in \citet{Howes11}, in an attempt to include nonlocal interactions, 
relevant for steep spectra.

In this letter we revisit KAW phenomenology, including in 
the energy transfer time 
the effect  of  ion temperature 
homogenization along magnetic field lines induced by Landau damping. 
The latter, whose linear rate is assumed
to persist in the nonlinear regime (see \citet{Schekochihin15} for a discussion),
not only provides the exponential cut-off 
(at a scale depending on the Kolmogorov constant) discussed in \citet{Howes06}, 
but also leads to a steepening  of  the $-7/3$ power law.
Furthermore, as it does not necessarily enforce strict 
critical balance, this model describes the transition between weak turbulence 
at large scales to strong turbulence at smaller scales. 

\section{Model setting}
We consider a  collisionless proton-electron plasma permeated by a strong 
ambient magnetic field (of
amplitude $B_0$), with  equal and isotropic mean temperatures $T_e=T_i$.
Alfv\'en waves are driven in the MHD range, at scales much larger than $\rho_i$.
The transverse magnetic field fluctuations $\delta B_\perp$ are measured in velocity units 
by defining $b= v_A (\delta B_\perp/B_0)$, where $v_A$ is the Alfv\'en velocity. 
The amplitude of these 
fluctuations at a transverse wavenumber $k_\perp$ is 
$b_k\sim (k_\perp E_k)^{1/2}$, where $E_k$ is the transverse magnetic spectrum. 

\subsection{Involved time scales}
The magnetic field being stretched by electron velocity gradients and the dynamics
dominantly transverse, we define, when the interactions are local,
a stretching frequency (inverse of the nonlinear time $\tau_{NL}$)
$\omega_{NL} \sim k_\perp v_{ek}$ where the transverse electron velocity $v_{ek}$ 
at scale $k_\perp^{-1}$ is given by 
$v_{ek}= {\overline \alpha} b_k$. Here, ${\overline \alpha}$ is a function of $k_\perp \rho_i$,  
equal to $1$ in the MHD range and scaling like $k_\perp \rho_i$ 
in the far sub-ion range, with a smooth transition near the ion gyroscale.
As in hydrodynamic turbulence, we write
$\omega_{NL} \sim [{\overline \alpha}^2 k_\perp^3 E_k]^{1/2}$ \citep{Kovasznay, Panchev69}.
Nevertheless, when $E_k$ is decaying fast enough, the above expression does not
necessarily ensure  the expected monotonic growth of $\omega_{NL}$.
In this case
nonlocal interactions cannot be neglected, and $\omega_{NL}$ should rather be viewed as the
stretching rate due to all the scales larger than $k_\perp^{-1}$. It is then  taken equal to
the r.m.s. value $\omega_{NL} = \Lambda [\int_0^{k_\perp} {\overline \alpha}^2 \,p_\perp^2 E_p dp]^{1/2}$
\citep{Ellison62,Panchev,Lesieur}, where $\Lambda$ is a constant. 
The local approximation is recovered when the integral diverges at large $k_\perp$,
while the integral formula can be replaced by the equation 
${d\omega_{NL}^2}/{dk_\perp} = \Lambda^2 {\overline \alpha}^2 \,k_\perp^2 E_k$. 

Alfv\'en waves are characterized by a frequency 
$\omega_W = {\overline \omega} k_\| v_A$ and a dissipation rate 
$\gamma= {\overline \gamma} k_\| v_A$. Here ${\overline \omega}$ 
(equal to $1$ in the MHD range) and ${\overline \gamma}$ are functions of 
$k_\perp \rho_i$ provided by the kinetic theory. In
a linear description, parallel and perpendicular wavenumbers are defined relatively to the ambient 
field (taken in the $z$-direction), but distortions of the magnetic field lines 
should be  retained in the nonlinear regime. 
We write  $k_\| = k_z + v_A^{-1}[\int_0^{k_\perp} p_\perp^2 E_p dp]^{1/2}$, where
$k_z$ is a constant estimating  the parallel
wavenumber of the energy-containing  Alfv\'en waves. It should not be confused with the inverse 
correlation length along the $z$-direction. Differently, $k_\|^{-1}$ measures the correlation
along the magnetic field lines (which is larger than along the $z$-axis).
A procedure to estimate $k_\|$ in numerical simulations is described in \citet{CL04},
while an approach using the frequency as a proxy for $k_\|$ is used in \citet{TenBarge12}.
This suggests to write 
$\omega_W = {\overline \omega} k_z v_A + \delta \omega_W$
with a turbulent frequency shift 
$\delta \omega_W = [\int_0^{k_\perp} {\overline \omega}^2 p^2 E_p dp]^{1/2}$ 
or $d{(\delta \omega_W)^2}/{dk_\perp} = {\overline \omega}^2 k^2 E_k$. 
We similarly write
$\gamma = {\overline \gamma} k_z v_A+ \delta \gamma$ with 
$d(\delta \gamma)^2/dk_\perp = {\overline \gamma}^2 k_\perp^2 E_k$. A  multiplicative constant
should also enter the definitions of $\delta\omega_W$ and $\gamma$, but is easily scaled out. 

Another time scale originates from the compressible character of KAWs. The latter are subject
to Landau damping, resulting in temperature
homogenization along the magnetic field lines, on the correlation length $k_\|^{-1}$ 
in a time $\tau_{H\,r} \sim (v_{th\, r} k_\|)^{-1}$.
Here the thermal velocity $v_{th\, r}$ appears as the r.m.s. streaming velocity  
of the $r$-particles. This time scale, which arises explicitly in 
Landau fluid closures \citep{HDP92,SHD97,SP15},
 is very short for the electrons, while for the
ions it is comparable to that of the  
other relevant processes and can thus affect the dynamics. 
Due to magnetic field distortion, this process introduces additional nonlinear couplings
characterized by the frequency 
$\omega_{H} = \mu v_{th\, i}k_\|$ where $\mu$ is a numerical constant. 
We are thus led to write $\omega_{H} = \mu v_{th \, i} k_z + \delta \omega_{H}$ with 
$d(\delta \omega_{H})^2/dk_\perp = \mu^2 \beta k_\perp^2 E_k$. 

Finally, we define the transfer time 
$\tau_{tr} = \tau_{NL} \left ({\tau_{NL}}/{\tau_{W}}+ {\tau_{NL}}/{\tau_{H}} \right)$,
or in frequency terms
$\omega_{tr} = {\omega_{NL}^2}/(\omega_{W} + \omega_{H})$.
When only one process competes with nonlinear stretching,
the critical balance condition
\citep{GS95,NS11,Scheko09} ensures the equality of the two associated time scales, 
and thus of the transfer and stretching times.

\subsection{The nonlocal model}
Retaining KAW linear Landau damping leads to the phenomenological equation  \citep{Howes08,Howes11}
$\partial_t E_k + {\mathcal T}_k = -2\gamma E_k + S_k$,
where $S_k$ is the driving term acting at large scales and ${\mathcal T}_k$ the transfer term 
related to the energy flux  $\epsilon$
by ${\mathcal T}_k = \partial \epsilon/\partial {k_\perp}$. Due to Landau damping, 
energy is not transferred conservatively along the cascade, making  $\epsilon$ 
scale-dependent. For  a steady state and outside the injection range, one has
${d\epsilon}/{dk_\perp} = - 2 \gamma E_k$. 
Note that the present setting differs from the asymptotic regime considered by \citet{Scheko09}
for which, under
the simultaneous conditions  $k_\perp \rho_i \gg 1$ and $k_\perp \rho_e \ll 1$,
KAWs are not subject to Landau damping, as they
transfer part of their energy via parallel phase mixing to the ion (electron) 
entropy cascades only at ion (electron) gyroscales, where it is cascaded both in physical 
and velocity spaces to collisional scales via perpendicular phase mixing.

Estimate of the energy flux relies on a basic turbulence  description
(overlooking intermittency). We write
$\epsilon = C \omega_{tr} k_\perp E_k$,   
where $C$ is a negative power of the Kolmogorov constant. 
Arguing that the small-scale eddies cannot be sufficiently highly correlated 
with each other to contribute equally \citep{Ellison62}, 
we here use the local approximation $k_\perp E_k$ of the Reynolds stress rather than 
the original  Obukhov's expression $\int_{k_\perp}^\infty E_p dp$ which leads to an
unphysical behavior in the dissipation range of hydrodynamic turbulence \citep{Panchev}.

Normalizing frequencies by $\Omega_i$, wavenumbers by $\rho_i^{-1}$,
energy spectra by $v_A^2 \rho_i$, energy fluxes by $v_A^2\Omega_i$, and denoting by $\beta$
the ion beta, we obtain the non-dimensional model
equations (keeping the same notations)
\begin{eqnarray}
&& {d\omega_{NL}^2}/{dk_\perp} = {\Lambda^2}{\beta}^{-1}{\overline \alpha}^2k_\perp^2 E_k \label{OMGNL}\\
&& {d(\delta \omega_W)^2}/{dk_\perp} = {\beta}^{-1} {\overline \omega}^2k_\perp^2 E_k\label{OMGW}\\
&& {d(\delta \gamma)^2}/{dk_\perp} =  {\beta}^{-1}{\overline \gamma}^2k_\perp^2 E_k \label{OMGG}\\
&& {d(\delta \omega_{H})^2}/{dk_\perp} = \mu^2k_\perp^2 E_k \label{OMGH}\\
&& {d\epsilon}/{dk_\perp} = -2 [\beta^{-1/2}{\overline \gamma}k_z + (\delta\gamma)]E_k \label{eps}\\
&&E_k =  [(\beta^{-1/2} {\overline \omega} + \mu)k_z + \delta\omega_W + \delta \omega_{H}] 
\frac{C^{-1}\epsilon}{k_\perp\omega_{NL}^2}.\nonumber \\
&&\label{spct}
\end{eqnarray}
Except possibly near $k_\perp=1$, ${\overline \alpha} = {\overline \omega}$.
The nonlinearity parameter $\chi = \omega_{NL}/ \omega_W$ obeys
\begin{equation}
\frac{d\chi}{dk_\perp} =\frac{{\overline \alpha}^2 k_\perp^2 E_k}{2\beta \omega_{NL}^2} \chi
\left (\Lambda^2 - \frac{{\overline \omega}^2}{{\overline \alpha}^2} \chi^2 \right ) -k_z 
\frac{\chi^3}{2\omega_{NL}^2}f
\end{equation} 
where 
$f= \frac{2}{\sqrt{\beta}} \frac{d}{dk_\perp}({\overline \omega}{\delta \omega_W})
+ \frac{k_z}{\beta} \frac{d{\overline \omega}^2}{dk_\perp}$ is positive.
At the (small) injection wavenumber $k_0$, turbulence is characterized by 
$A= k_z / [k_0^3 E_{k_0}]^{1/2}= (k_z/k_0)(B_0/\delta B_{\perp 0})$ and,
when taking  $\omega_{NL}^{(0)} = \Lambda \beta^{-1/2}  k_0^{3/2} E_{k_0}^{1/2}$ 
and $\omega_W^{(0)} = \beta^{-1/2} (k_z + k_0^{3/2} E_{k_0}^{1/2})$, 
$\chi_0 = \Lambda /(1+A)$. In the strong turbulence regime with $k_z =0$,
$\chi= \Lambda$ in the full spectral range, while for $k_z \neq 0$,  
$\chi$  starts growing near $k_0$ but cannot exceed $\Lambda$. 

\section{A simplified local-interaction model}
\begin{table*}
\begin{center}
\def~{\hphantom{0}}
\begin{tabular}{|c||c|c|c|c|c|c|c|}
\hline
$\Lambda$ & $0.71$   & $1$    & $1.22$  & $1.30$  & $1.41$   & $2$     & $4.47$ \\
\hline
exponent  & $-3.18$  & $-2.81$ & $-2.69$ & $-2.66$ & $-2.63$ & $-2.53$ & $-2.45$\\
\hline
\end{tabular}
\caption{Sub-ion exponent versus $\Lambda$ in conditions of Fig. \ref{F1}a 
(fitting range $8\le k_\perp \le 40$).}
\end{center}
\end{table*}

\subsection{The usual conservative cascades}

Assuming local interactions, we have
\begin{equation}
E_k^2 \sim  \frac{\beta^{1/2} \Lambda^{-2}C^{-1}\epsilon}{{\overline \alpha}^2 k_\perp^4 }({\overline \omega}
+ \mu\beta^{1/2})(k_z + k_\perp^{3/2} E_k^{1/2}).
\end{equation}
When neglecting dissipation (constant $\epsilon$), we recover the usual inertial
energy spectra. For  weak turbulence
($k_\perp^{3/2}E_k^{1/2}  \ll k_z$),  $E_k \propto k_\perp^{-2}$  
in the MHD range, while in the 
sub-ion range (as ${\overline \omega}\gg \mu\beta^{1/2}$), 
$E_k \propto k_\perp^{-5/2}$. For strong turbulence
($k_z$ negligible), $E_k \propto k_\perp^{-5/3}$ in the MHD range, while in the sub-ion range 
$E_k \propto k_\perp^{-7/3}$. A $k_\perp^{-3}$ regime is also obtained in the sub-ion range 
when the effect of wave propagation is negligible, as observed in 
two-dimensional hybrid PIC simulations with an out-of-plane ambient magnetic field \citep{Franci}.
Furthermore, as easily seen from eq. (\ref{spct}), an additional regime with  
$E_k \propto k_\perp^{-1}$ is possible at very large scales when, for small enough $\epsilon$, 
turbulence is not yet developed and $\omega_{NL}$ almost constant. 
Such a spectral exponent is observed in the solar wind 
at scales larger than the $k_\perp^{-5/3}$ inertial range 
\citep{Matthaeus86,Nicol08,Wicks10}.

\begin{figure}
\begin{center}
\includegraphics[width=0.48\textwidth]{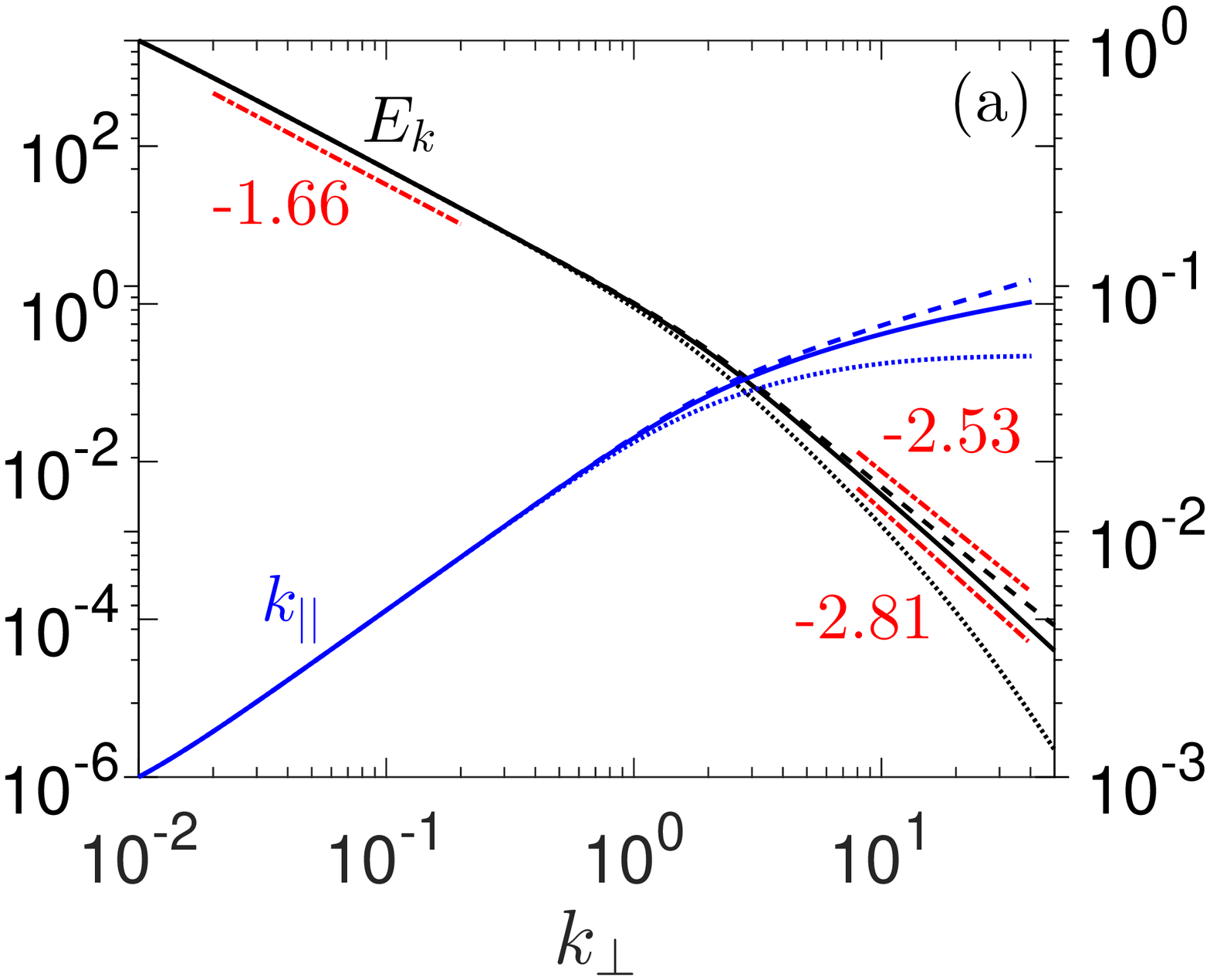}
\includegraphics[width=0.48\textwidth]{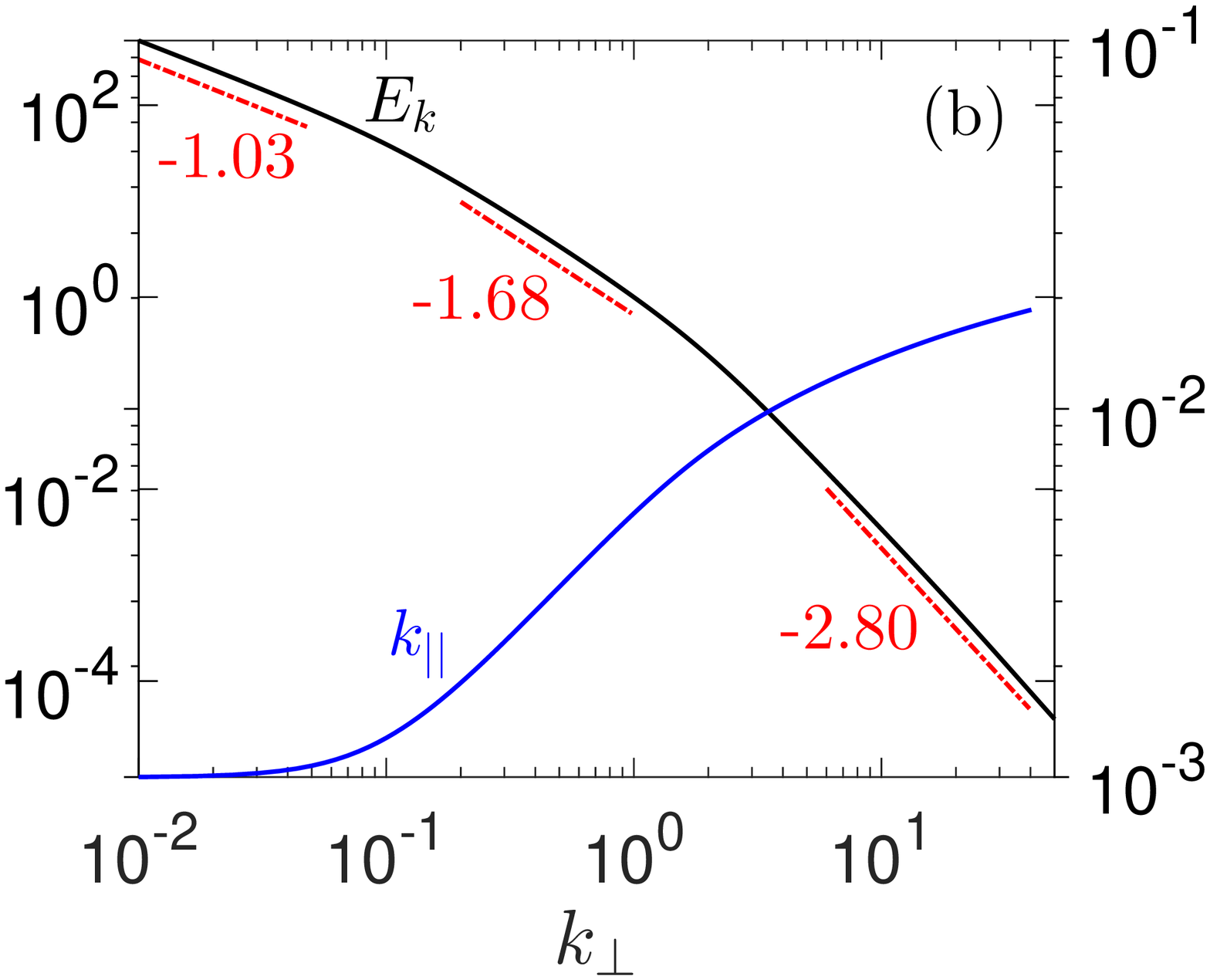}
\end{center}
\caption{(a): Normalized energy spectrum $E_k$ 
(black lines, l.h.s. labels and outer tickmarks) 
and parallel wavenumber $k_\|$ (blue lines, r.h.s labels and inner tickmarks), for 
$k_z=0$, $\beta=e=1$ and 
$\Lambda= 2$ (dashed lines), $1$ (solid lines) and $0.5$ (dotted lines),  
(see other parameters in text); (b): Same as (a) for
$\Lambda=1$ and $\epsilon_0=10^{-2}$.
}
\label{F1}
\end{figure}

\subsection{Effect of Landau damping}
For strong turbulence, when assuming local interactions and neglecting $k_z$ contributions,
\begin{equation}
\epsilon_k = \epsilon_0 \exp \left[-2C^{-1}\Lambda^{-2} \int_{k_0}^{k_{\perp}} 
\frac{{\overline \gamma}}{\xi {\overline \alpha}^2}
({\overline \omega} + \mu\beta^{\frac{1}{2}})d\xi\right], \label{epsilonk}
\end{equation}
where the notation $\epsilon_k$ for the energy flux stresses its wavenumber dependence.
This quantity is to be substituted in 
$E_k \sim \beta^{1/3} \Lambda^{-4/3} C^{-2/3} \epsilon_k^{2/3} k_\perp^{-7/3}$, taking 
${\overline \alpha} = {\overline \omega}$ and 
${\overline \gamma}/{\overline \omega}^2 \approx \delta(\beta)$
($\approx 0.78 \rho_e/\rho_i$
when using eqs. (62) and (63) of \citet{Howes06} with $\beta =1$). Furthermore, 
in this range ${\overline \omega} \approx a(\beta) k_\perp$ with  $a(\beta) = (1+\beta)^{-\frac{1}{2}}$.
This leads to $\epsilon_k \sim \epsilon_0 k_\perp^{-\zeta} 
\exp (-2 a(\beta) C^{-2}\Lambda^{-2}\delta(\beta) k_\perp)$, 
with $\zeta= 2\delta(\beta) C^{-1}\mu\Lambda^{-2} \beta^{1/2}$.
This results in  a steepening of the algebraic prefactor of the magnetic spectrum 
which is now proportional to 
$k_\perp^{-(7/3 + 2\zeta/3)} \exp [-(4/3) a(\beta)\delta(\beta) C^{-1}\Lambda^{-2} k_\perp]$.
Compared with the spectrum obtained in \cite{Howes08}, the present model predicts that, 
in addition to an  exponential cut-off, Landau damping leads to a  correction of 
the power-law exponent. In contrast with intermittency corrections discussed in \citet{Boldyrev12},
this correction is not universal, which is expected when dissipation
and nonlinear transfer times display the same wavenumber dependence \citep{Bratanov13}.
This situation holds for the second term in the exponential arising in  eq. (\ref{epsilonk}), that leads
to the correction of  the $-7/3$ spectral index. We note that it defines a dissipation length
$k_d^{-1}= 2 C^{-1} \Lambda^{-2}\mu \beta^{1/2} {\overline \gamma}/(k_\perp {\overline \alpha^2}) \propto k_\perp^{-1}$ 
and thus a dissipation rate $\omega_d = v_e k_d$ for $\epsilon_k$, which has the same $k_\perp$
dependency as $\omega_{NL}$ that, by a critical balance argument, identifies with $\omega_{tr}$
beyond the transition range. A similar power law decay of $\epsilon$ is encountered in 
drift-kinetic plasma turbulence (see eq. (2.53) of \cite{Schekochihin15}).

Differently, for weak turbulence, we get
\begin{equation}
\epsilon_k = \left [\epsilon_0^{\frac{1}{2}} - 
\Lambda^{-1} C^{-\frac{1}{2}}\beta^{-\frac{1}{4}} k_z^{\frac{3}{2}} \int_{k_0}^{k_\perp} 
\frac{{\overline \gamma}}{\xi^2{\overline \alpha}^{\frac{1}{2}}} d\xi\right]^2, \label{eps-weak}
\end{equation}
where the integral behaves like $k_\perp^{1/2}$. Equation (\ref{eps-weak}) predicts that
$\epsilon_k$ and thus $E_k$ vanish at a finite $k_\perp$, indicating the breaking of
the analysis near the corresponding scale, an effect possibly related to the difficulty 
for weak turbulence to exist in the presence of a significant Landau damping.

\begin{figure}
\begin{center}
\includegraphics[width=0.48\textwidth]{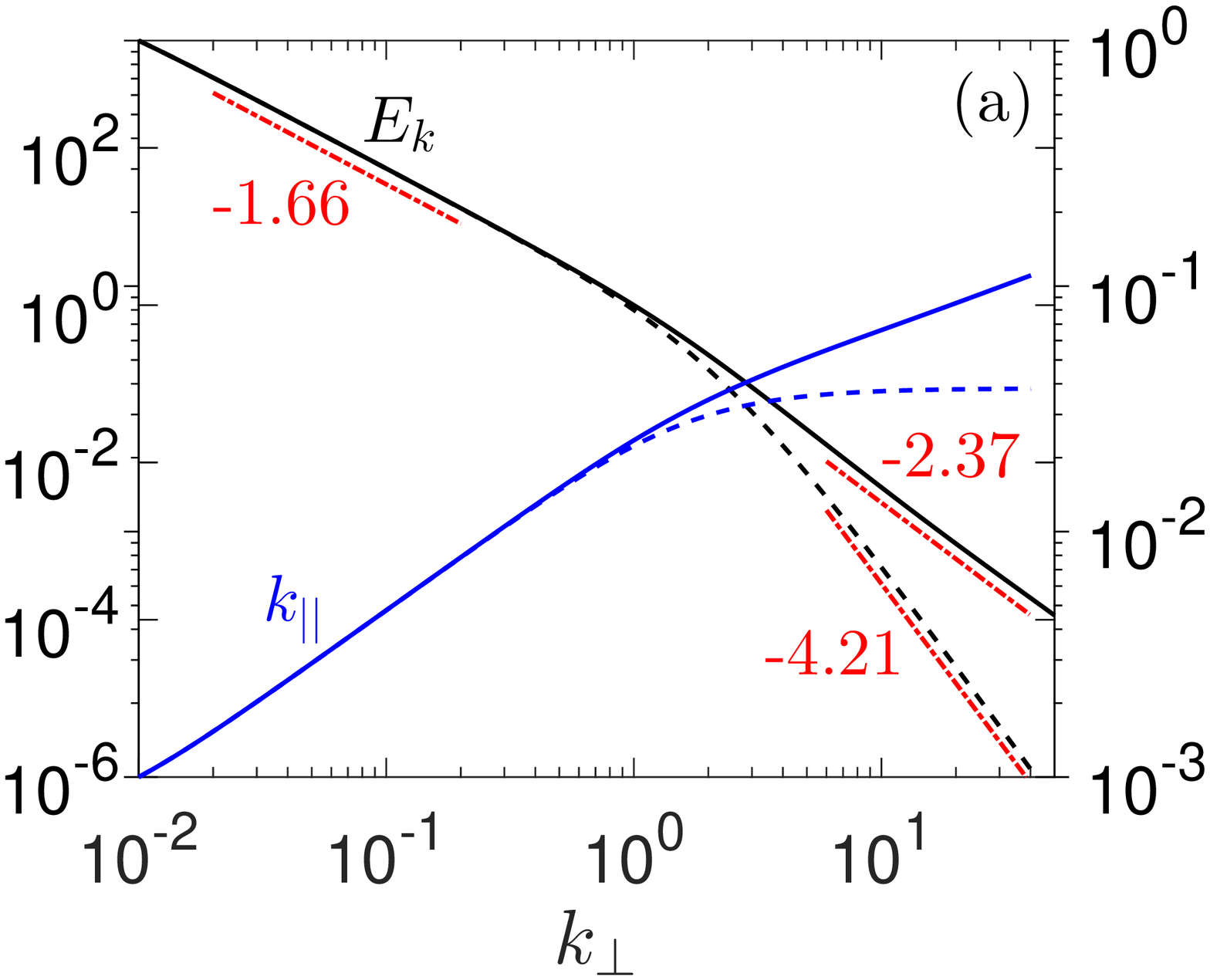}
\includegraphics[width=0.48\textwidth]{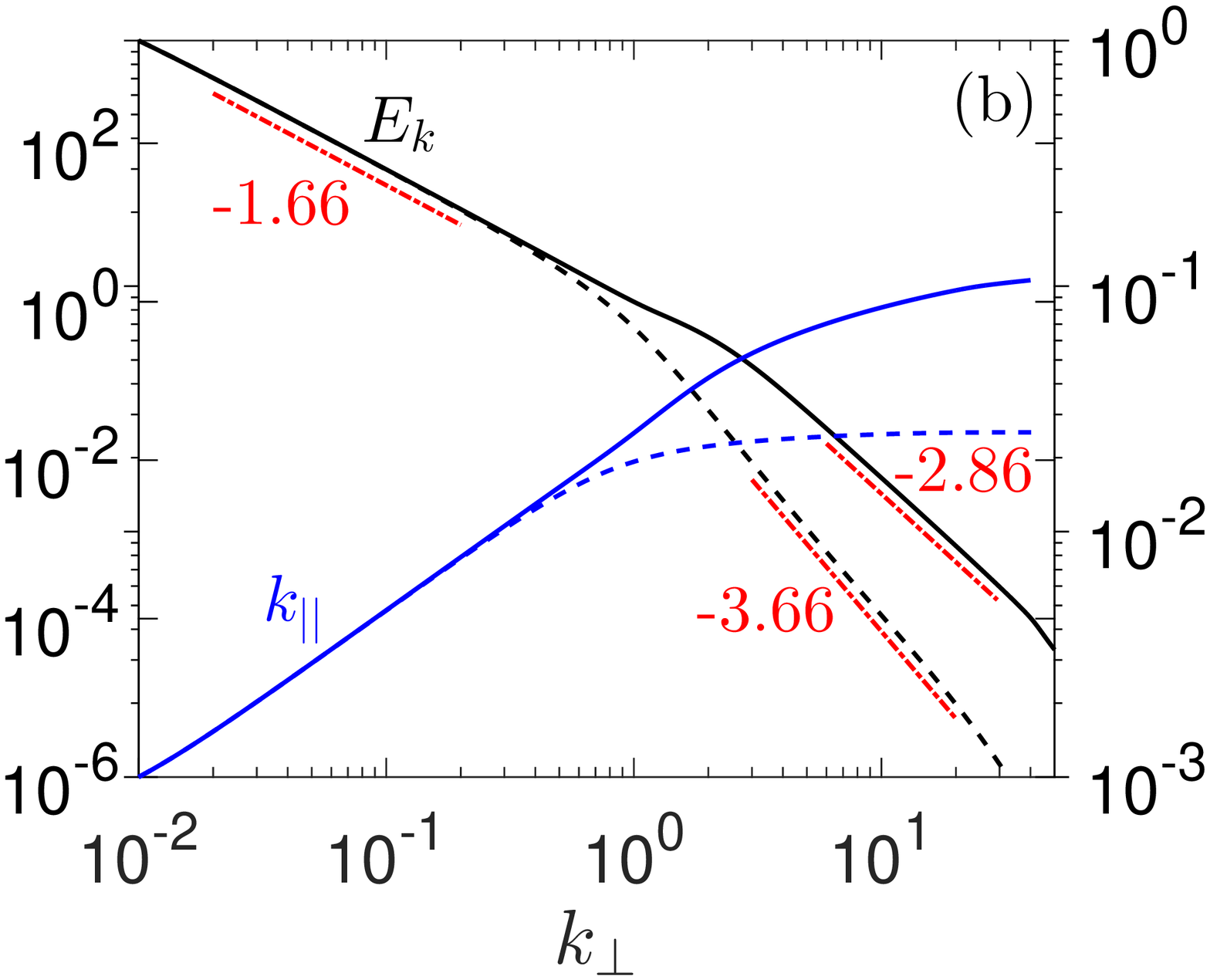}
\end{center}
\caption{Same as Fig. 1a  with  $\beta=0.01$,  $\Lambda= 1$ (solid lines)and $0.45$ 
(dashed lines) (a), and $\beta=10$,
$\Lambda=3.16$ (solid lines) and $1$ (dashed lines)  (b).}
\label{F2}
\end{figure}

\section{Numerical integration of the full model}

When the spectrum is too steep, a numerical integration of differential equations  
(\ref{OMGNL})-(\ref{spct})  is needed. 
The functions ${\overline \omega}$ and ${\overline \gamma}$ are then
evaluated from the full linear kinetic theory by means of the  WHAMP software \citep{WHAMP}.
Moreover, ${\overline \alpha} = {\overline \omega}$.
We prescribed conditions at $k_\perp =k_0$
in the form $\omega_{NL}^{(0)} = \Lambda \beta^{-1/2} {\overline \alpha_{k_0}} k_0^{3/2} E_0^{1/2}$, 
$\delta\omega_W^{(0)} = \beta^{-1/2} {\overline \omega_{k_0}} k_0^{3/2} E_0^{1/2}$, 
$\delta\gamma^{(0)} = \beta^{-1/2} {\overline \gamma_{k_0}} k_0^{3/2} E_0^{1/2}$,
$\delta\omega_{H}^{(0)} = \mu k_0^{3/2} E_0^{1/2}$ and 
$\epsilon_0= e C k_0\omega_{NL}^{(0)2} E_0/[(\beta^{-1/2} {\overline \omega} 
+ \mu)k_z + \delta\omega_W + \delta \omega_{H}]$. Here $E_0$ is an arbitrary constant
(taken equal to $1$ with no lack of generality). We chose $k_0=10^{-2}$,
$C=1.25$ and $\mu=1.8$. Except when otherwise specified, we also took $e=1$. 
For clarity's sake, when several energy spectra are plotted in the same panel, 
one of them  (solid line)
is normalized by its value at $k_\perp =1$, while  the others  (dashed and doted lines) are rescaled to 
make all the spectra equal at  $k_\perp = k_0$.
In red are indicated the fitting ranges (dashed-dotted straight lines)
and the corresponding spectral exponents whose last digit only is sensitive to
moderate changes of the fitting domain.

In Fig. \ref{F1}a, we focus on the strong turbulence regime for  $\beta=1$,
assuming $k_z=0$. In the MHD range, a $k_\perp^{-5/3}$ 
energy spectrum establishes and  $k_\|$  scales
like $k_\perp^{2/3}$ in all the cases. Differently, at the sub-ion scales, the spectrum is steeper when
$\Lambda$ is smaller, displaying exponents $-2.53$ for $\Lambda= 2$, $-2.81$ for 
$\Lambda=1$, and a fast decay for $\Lambda=0.5$.
In this range, $k_\|$ increases slower, and this even more so when $\Lambda$ 
is smaller. In the present setting, 
$\Lambda\approx 0.71$ (leading to a spectral exponent  $-3.18$) and
$\Lambda\approx 4.47$ (exponent  $-2.45$) appear as the extreme values for existence of
an extended  power-law spectrum at the sub-ion scales.
These limit exponents are consistent with the 
dispersion of solar-wind measurements.
Exponents for intermediate values of $\Lambda$ are 
displayed in Table 1.
Their statistical distribution (and the most probable value) are
nevertheless beyond the scope of the model. Note that a
$-7/3$ exponent (rarely reported in observations)
is approached for large $\Lambda$, only when $\mu=0$.

When keeping $\Lambda = 1$ but decreasing the energy transfer rate 
$\epsilon_0$ by taking $e=10^{-2}$ (Fig. \ref{F1}b), the MHD and sub-ion spectral
exponents are not affected, but a  $k_\perp^{-1}$ range becomes
visible at the largest scales. In this range, where $k_\|$ remains small,
turbulence is not developed,
possibly as in the solar wind energy containing range. Further decrease of $\epsilon_0$
leads to a $k_\perp^{-1}$ range extending down to the ion gyroscale,
a situation observed in the magnetosheath near the bow shock \citep{Czayk01,Alexandrova08}.
 
Another issue is the influence of $\beta$, keeping $k_z=0$.
For $\beta=0.01$ (Fig. \ref{F2}a), we considered $\Lambda=1$ and $\Lambda = 0.45$,
leading to sub-ion spectral exponents $-2.37$ and $-4.21$ respectively,
while as expected the MHD range is not affected.
For $\beta=10$ (Fig. \ref{F2}b), we used $\Lambda=3.16$ and $1$, for which 
the sub-ion exponents are $-2.86$ and $-3.66$ respectively. 
For the former value of $\Lambda$, a spectral bump is visible
in the transition zone, consequence of the sharp decrease of the 
ion Landau damping at these scales (see e.g. Fig. 3 of \citet{Howes06}). 

To justify the chosen values of $\Lambda$ (which is equal to $\chi$ in the above setting 
where $k_z =0$ and prescribes its saturated value when $k_z\ne 0$), 
it should be noted that, at a fixed wavenumber,
$\chi$ increases with the amplitude of the fluctuations,
linearly in weak turbulence and at a slower rate
when the amplitude gets larger, a behavior supported by 
FLR-Landau fluid simulations  (in preparation). Furthermore,
the Mach number, which scales like  $b_0 \beta^{-1/2}$ (where
$b_0$ measures the amplitude of the large-scale turbulent fluctuations) is
usually observed to be moderate in the solar wind \citep{Bavassano95}, in 
spite of the broad range of reported values of $\beta$ \citep{Chen14}. This
implies that the turbulence level, and thus $\Lambda$, should be decreased
at smaller $\beta$.

\begin{figure}
\begin{center}
\includegraphics[width=0.48\textwidth]{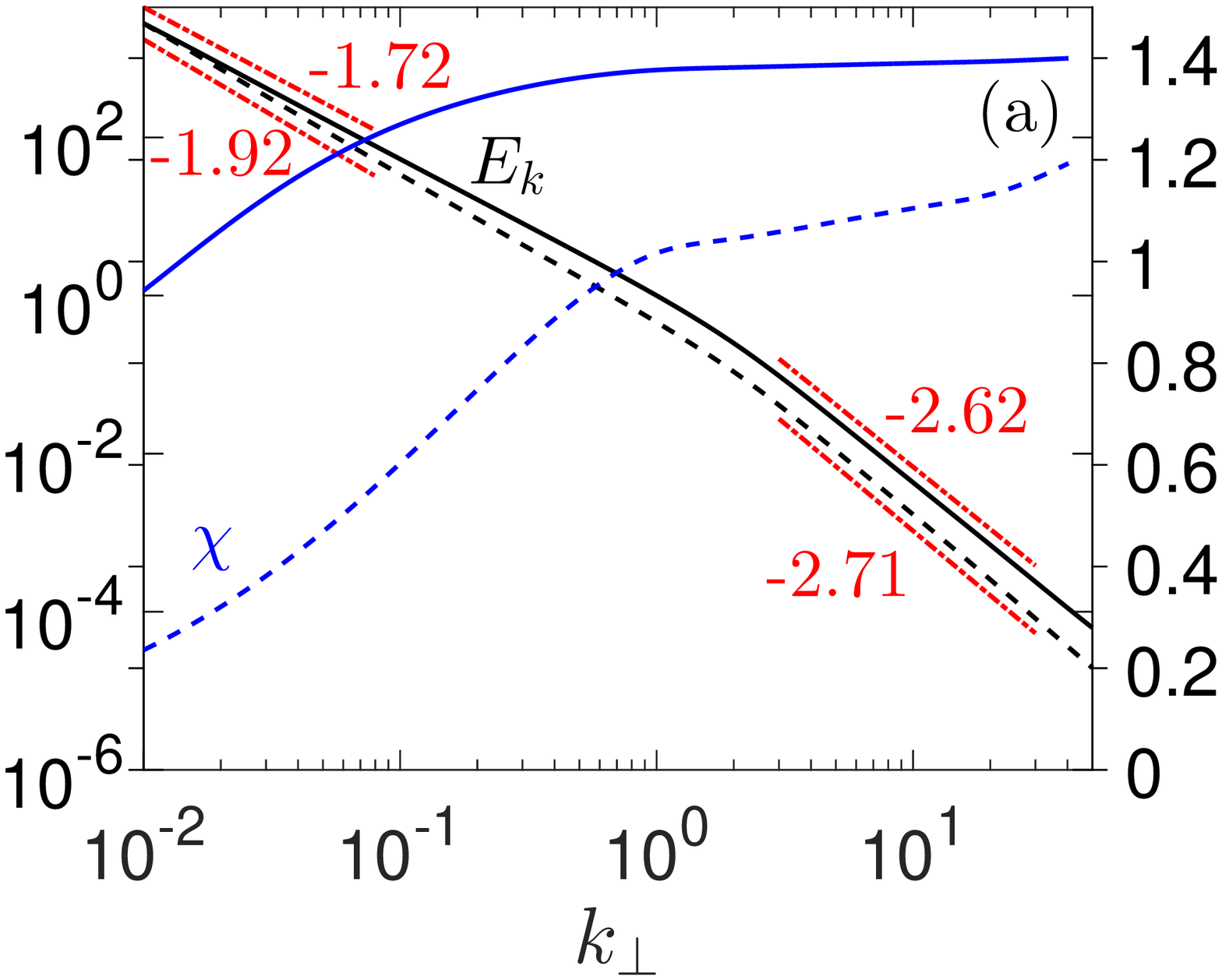}
\includegraphics[width=0.48\textwidth]{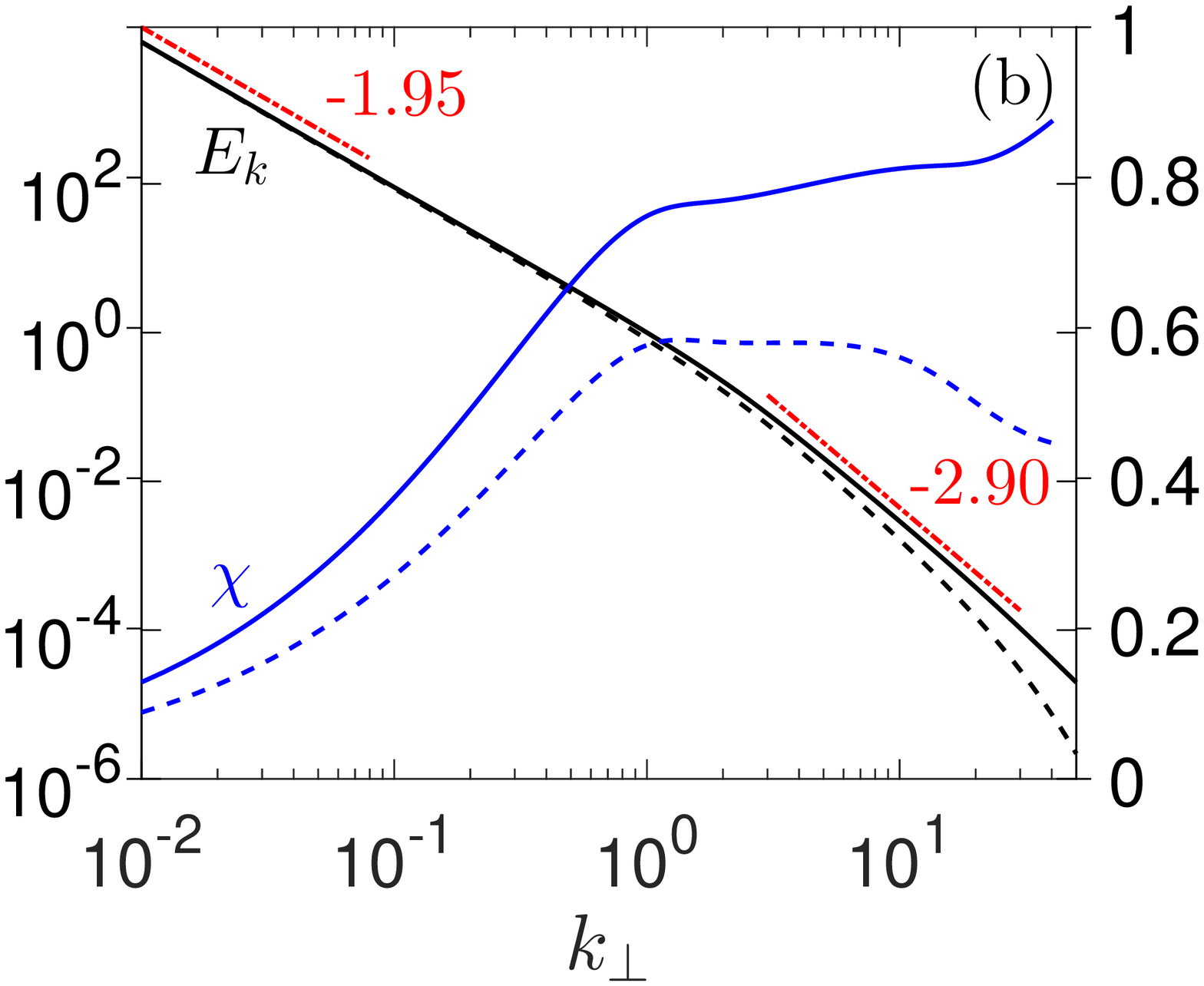}
\end{center}
\caption{(a): Energy spectrum $E_k$ (l.h.s. labels) and nonlinear parameter $\chi$ 
(r.h.s. labels) for 
$\Lambda = \sqrt{2}$ with  $A=0.5$ (solid lines) and $A=5$ (dashed lines);
(b): Same as (a),  with $A=10$ (solid lines) and $A=15$ (dashed lines).
}
\label{F3}
\end{figure}

Figure \ref{F3} addresses the influence of the parameter $A$ when $\Lambda$
is fixed at a moderate value, here $\sqrt{2}$. Increasing $A$
results in a change from  strong to weak 
turbulence near the driving scale. 
For $A=0.5$ (Fig. \ref{F3}a), the function $\chi$ rapidly saturates to a value
slightly smaller than $\Lambda$, establishing critical balance and thus 
a strong turbulence regime. The spectral exponent $-1.72$ measured in the 
MHD range differs from  $-5/3$, since $\chi$ is still in the
growing phase, but choosing $k_0$ smaller would ensure a $k_\perp^{-5/3}$ critically-balanced 
MHD range. 
At small scales, the exponent is $-2.62$, comparable to the values
displayed in Fig. \ref{F1}a. The case $A=5$ corresponds to an intermediate regime where $\chi_0$
is significantly smaller, resulting in a  $-1.92$ large-scale spectrum, 
steeper than $-5/3$ but nevertheless distinguishable from the $-2$ weak-turbulence value. 
The sub-ion exponent  $-2.71$ is consistent with a strong turbulence regime,
characterized by an almost constant $\chi$ at these scales.
In Fig.  \ref{F3}b,  $A = 10$ and $15$, both lead to a weak-turbulence regime at large
scales, with exponents $-1.95$ and $-1.97$, very close to the theoretical value.
Differences nevertheless hold  at
small scales. For $A=10$, the sub-ion dynamics corresponds to strong turbulence,
qualitatively similar to the case $A=5$,
with a spectral exponent $-2.9$ and an almost constant  $\chi$, although somewhat smaller.
Differently, for $A=15$,  the fluctuations are
too weak for the energy transfer to efficiently  compete
with Landau damping, leading to an exponential decay of sub-ion spectrum, and a function 
$\chi$ which starts to decrease  by $k_\perp \rho_i=10$.

\section{Conclusion} 

This model predicts
a non-universal power-law spectrum  for strong turbulence at the
sub-ion scales  (beyond the transition range) with  an exponent which,
in contrast with the $-5/3$ inertial MHD cascade, depends  on the saturation
level of the nonlinearity parameter $\chi$, covering a range of values
consistent with solar wind and magnetosheath observations.
Such a non universality, associated with Landau damping, was also reported 
in three-dimensional PIC simulations of whistler turbulence \citep{Gary12}
and FLR-Landau fluid simulations of KAW turbulence (in preparation). 
As the present approach does not capture sub-electron scale dynamics,
the exponential cut-off might be replaced by another regime, such as the 
steep power laws reported from both numerical simulations \citep{Campo11,Gary12} 
and spacecraft data \citep{Sah13}.

\acknowledgments 
The research leading to these results has received funding from the 
European Commission's Seventh Framework Programme (FP7/2007-2013) under
the grant agreement SHOCK (project number 284515). 


\end{document}